\documentclass[runningheads]{llncs}
\usepackage{graphicx}
\begin{document}
\title{ITM-Rec: An Open Data Set for Educational Recommender Systems}
\author{Yong Zheng}
\authorrunning{Yong Zheng}
\institute{Department of Information Technology \& Management\\
Illinois Institute of Technology, Chicago, IL 60661, USA\\
\email{yzheng66@iit.edu}
}
\maketitle              % typeset the header of the contribution
\begin{abstract}
With the development of recommender systems (RS), several promising systems have emerged, such as context-aware RS, multi-criteria RS, and group RS. However, the education domain may not benefit from these developments due to missing information, such as contexts and multiple criteria, in educational data sets. In this paper, we announce and release an open data set for educational recommender systems. This data set includes not only traditional rating entries, but also enriched information, e.g., contexts, user preferences in multiple criteria, group compositions and preferences, etc. It provides a testbed and enables more opportunities to develop and examine various educational recommender systems.

\keywords{recommender system \and data set \and context-aware \and group \and multi-criteria \and ITM-Rec\footnote{This paper was accepted and published in the Companion Proceedings 13th International Conference on Learning Analytics \& Knowledge (LAK), 2023, https://www.solaresearch.org/publications/conference-proceedings/}}
\end{abstract}
\section{Introduction}Nowadays, recommender systems (RS) have become one of the major tools in technology-enhanced learning. RS have been introduced to the education domain~\cite{santos2011educational} to deliver personalized learning, recommend textbooks or informal learning materials, assist decision-making in group studies or teamwork, and provide better adaptive learning in mobile environments, etc.

Various RS were well-developed with enriched information, such as contextual variables and multi-criteria user preferences. However, the education domain may not benefit from these RS due to a lack of corresponding data sets with necessary information. In this paper, we fill this gap by announcing and releasing the ITM-Rec data set, which can be utilized as a testbed to develop and examine different types of RS~\cite{zheng2018cars,zheng2022multi,zheng2022survey,zheng2019educational} in educations.

\section{ITM-Rec: Data Collections, Statistics \& Usage}

\subsection{Data Collections}
The data set was collected from a questionnaire using Qualtrics from 2017~\cite{zheng2018personality}. The subjects were graduate students enrolled in the specialization of data management and analytics at the ITM department in Illinois Tech. The questionnaire was designed to collect student preferences on the topics of the final projects in three courses: database (DB), data analytics (DA), and data science (DS).

More specifically, students in the DA and DS classes were given a list of Kaggle data sets as candidates for analysis in their final projects. Each student was asked to select at least three data sets they liked and three that they did not like. They were asked to rate their selections by giving an overall rating and additional ratings for three criteria: App (their liking of the application domain), Data (their liking of the data's processing or storage), and Ease (their liking of the degree of ease in using the data for the final project). In the DB class, students were asked to build web information systems with connections to relational databases, and were asked to rate different potential project topics (e.g., hotel booking systems, hospital appointment systems) by giving overall ratings and the same multi-criteria ratings. The questionnaire was first assigned to each student to collect individual preferences. In addition to user and group preferences, we also collected student demographic information (e.g., age, gender, marriage status) and item content features (e.g., URL of Kaggle data, title of the data, textual descriptions of the data). 

Regarding the final projects, students had the option to work independently or as part of a team. If a team was formed for the final project, each group of students were asked to complete the questionnaire again as a second survey. It is worth noting that this second survey was assigned to each group and they were required to submit a single copy of the questionnaire after group discussions. Their input can be considered as group preferences on the items, rather than individual preferences.

\subsection{Data Description and Statistics}
We have collected the data from 2017 to 2022. The user information has been well anonymized, and data set will be released on Github\footnote{https://github.com/irecsys/RecData } and Kaggle\footnote{https://www.kaggle.com/datasets/irecsys/itmrec}. The whole data set is composed of five major files:
\begin{itemize}
\item Users.csv, where we provide meta data (ID, gender, age, etc.) about 476 unique students.
\item Items.csv, where we provide meta data (ID, title and descriptions) about 70 unique items.
\item Ratings.csv refers to individual preferences and it contains 5,230 ratings given by 476 users over 70 items. An example is shown in Table 1. In addition to overall and multi-criteria ratings, context information such as the course (DB, DA or DS), semester (Spring, Fall) and COVID-19 lockdown periods (PRE, DUR, POS) is provided. PRE refers to the timeframe from 2017 Fall to 2019 Fall, DUR refers to the timeframe from 2020 Spring to 2021 Spring, and the POS period refers to the timeframe from 2021 Fall to 2022 Fall. 
\item Group\_ratings.csv refers to group preferences and contains 1,117 ratings given by 143 groups.
\item Group.csv describes the composition of groups. There are 143 groups, where 88, 42, 9, 4 groups have a group size of 2, 3, 4, 5, respectively.
\end{itemize}

\begin{table}[ht!]
\centering
\caption{Example of Individual Ratings by Students}
\begin{tabular}{|c|c|c|c|c|c|c|c|c|}
\hline
\textit{UserID} & \textit{ItemID} & \textit{Rating} & \textit{App} & \textit{Data} & \textit{Ease} & \textit{Course} & \textit{Semester} & \textit{Lockdown} \\ \hline
1173            & 28              & 5               & 4            & 4             & 4             & DA              & Fall              & PRE               \\ \hline
1175            & 41              & 5               & 4            & 4             & 4             & DS              & Spring            & POS               \\ \hline
…               & …               & …               & …            & …             & …             & …               & …                 & …                 \\ \hline
\end{tabular}
\end{table}

\subsection{Data Usage}
Due to the enriched information, this data can be utilized as a testbed to develop and examine various recommender systems. Below are existing or possible examples by using this educational data set:
\begin{itemize}
\item Traditional RS with side information (e.g., user demographic information and item features), or identification of grey sheep users in RS~\cite{zheng2017identification}. 
\item Context-aware RS~\cite{zheng2018cars,zheng2014deviation} which adapt the recommendations to different contexts (e.g., course, semester, lockdown periods), or context suggestions~\cite{zheng2016user} which recommend contexts, rather than items.
\item  Multi-criteria RS which enhance recommendations by taking advantage of user preferences in multiple criteria (e.g., App, Data, Ease)~\cite{zheng2022multi}.
\item Group RS~\cite{zheng2018exploring} which recommend items to each group, rather than individuals.
\item Multi-objective RS~\cite{zheng2022survey} which optimize multiple objectives in RS.
\item Integrated RS with multi-factors, where researchers can build RS by integrating various factors together, e.g., integrating the context information in multi-criteria RS~\cite{zheng2019integrating}, utilizing multi-criteria preferences towards group RS~\cite{zheng2019educational}, etc. 
\end{itemize}

\section{Conclusions and Future Work}
In this paper, we announce and release an open data set for educational recommender systems, where enriched information (e.g., contexts, multi-criteria preferences, group compositions and preferences) are available. As a result, this data set can be utilized to develop and examine different types of educational recommender systems. We may continue the process of data collections, and release another version of this data in the next few years.

\bibliographystyle{splncs04}
\bibliography{sample-base}

\begin{thebibliography}{10}
\providecommand{\url}[1]{\texttt{#1}}
\providecommand{\urlprefix}{URL }
\providecommand{\doi}[1]{https://doi.org/#1}

\bibitem{santos2011educational}
Santos, O.C.: Educational recommender systems and technologies: Practices and
  challenges: Practices and challenges  (2011)

\bibitem{zheng2014deviation}
Zheng, Y.: Deviation-based and similarity-based contextual slim recommendation
  algorithms. In: Proceedings of the 8th ACM Conference on Recommender Systems.
  pp. 437--440 (2014)

\bibitem{zheng2018exploring}
Zheng, Y.: Exploring user roles in group recommendations: A learning approach.
  In: Adjunct Publication of the 26th Conference on User Modeling, Adaptation
  and Personalization. pp. 49--52 (2018)

\bibitem{zheng2018personality}
Zheng, Y.: Personality-aware decision making in educational learning. In:
  Proceedings of the 23rd international conference on intelligent user
  interfaces companion. pp.~1--2 (2018)

\bibitem{zheng2019educational}
Zheng, Y.: Educational group recommendations by learning group expectations.
  In: 2019 IEEE International Conference on Engineering, Technology and
  Education (TALE). pp.~1--7. IEEE (2019)

\bibitem{zheng2017identification}
Zheng, Y., Agnani, M., Singh, M.: Identification of grey sheep users by
  histogram intersection in recommender systems. In: Advanced Data Mining and
  Applications: 13th International Conference, ADMA 2017, Singapore, November
  5--6, 2017, Proceedings 13. pp. 148--161. Springer (2017)

\bibitem{zheng2018cars}
Zheng, Y., Mobasher, B.: Context-aware recommendations  (2018)

\bibitem{zheng2016user}
Zheng, Y., Mobasher, B., Burke, R.: User-oriented context suggestion. In:
  Proceedings of the 2016 Conference on User Modeling Adaptation and
  Personalization. pp. 249--258 (2016)

\bibitem{zheng2019integrating}
Zheng, Y., Shekhar, S., Jose, A.A., Rai, S.K.: Integrating context-awareness
  and multi-criteria decision making in educational learning. In: Proceedings
  of the 34th ACM/SIGAPP Symposium on Applied Computing. pp. 2453--2460 (2019)

\bibitem{zheng2022multi}
Zheng, Y., Wang, D.: Multi-criteria ranking: Next generation of multi-criteria
  recommendation framework. IEEE Access  \textbf{10},  90715--90725 (2022)

\bibitem{zheng2022survey}
Zheng, Y., Wang, D.X.: A survey of recommender systems with multi-objective
  optimization. Neurocomputing  \textbf{474},  141--153 (2022)

\end{thebibliography}

\end{document}